\documentstyle[12pt,epsf]{article}
\newcommand{\bm}{\bibitem}

\begin{document}
\normalbaselineskip = 24 true pt
\normalbaselines
\thispagestyle{empty}
\rightline{\large\sf SINP-TNP/96-15}
\rightline{\large Oct 1996}
\vskip 0.5 cm
\begin{center}
{\Large {Dilepton Yield in Heavy-Ion Collisions with
 Bose Enhancement of Decay Widths}}
\vskip 1.2 cm
{\bf Abhee K. Dutt- Mazumder$^{a,}${\footnote
{email: abhee@saha.ernet.in}}, Jan-e Alam$^b$,
Binayak Dutta-Roy$^a$\\ and\\ Bikash Sinha$^{a,b}$}\\[1cm] 
$^a$Saha Institute of Nuclear Physics\\
1/AF, Bidhan Nagar\\ Calcutta - 700 064, India\\[0.5cm]
$^b$Variable Energy Cyclotron Centre\\
1/AF, Bidhan Nagar\\ Calcutta - 700 064, India\\
\end{center}
\vskip 1.0 cm
\begin{abstract}
The excess of low invariant mass dilepton yield in heavy ion collisions
arising from reduction in the rho meson mass at finite temperatures is
partially suppressed 
because of the effect on the width of the rho meson induced by
Bose enhancement, essentially due to emission of pions
in a medium of the pion gas in the central 
rapidity region. The sensitivity of the effect on the initial temperature
of the hadronic phase is also examined. 
\end{abstract}
\newpage
Heavy ion collisions (HIC) provide an unique probe to study the properties
of hot and dense nuclear matter temporarily produced when two highly energetic
heavy nuclei collide in the laboratory.
Various particles are produced directly and via
secondary collisions in such high energy HIC 
which serve as a tool to extract information about the environment
leading to their production; among these the dileptons,
as well as the photons, seem to be rather promising, as they
interact only electromagnetically  with the rest of the matter and therefore, 
suffer minimum final state interaction. With this in mind several experiments
have been performed to measure the dilepton yield in HIC \cite{gale,xia}. 
This could also provide
important signals for the novel state of matter - the Quark Gluon
Plasma (QGP) phase, expected to be produced in collisions of two
nuclei at ultra-relativistic energies, as per conventional 
wisdom of Quantum Chromodynamics.

In reality, the contributing factors to this spectra are many and
may in fact come from the different stages of the evolution of the hot
and dense hadronic matter temporarily produced in the preliminary stage of
the reaction. In the hadronic stage major contributions may come,
say for example, from
$\pi\pi\rightarrow e^+e^-$, $\pi\rho\rightarrow \pi e^+e^-$,
$\omega\rightarrow\pi^0e^+e^-$, among many
others as discussed in Ref. \cite{dks96}. 
In this letter, 
we demonstrate the effect on low invariant  mass dilepton
yield in heavy ion collisions arising from the 
increase in width of vector mesons (decaying into a pion gas in the 
central rapidity region) due to Bose enhancement (BE) 
through the induced emission of pions. For the sake of illustration
and concreteness we study in particular the role of the 
thermal nucleon loop in the rho-mediated 
production of dileptons, more precisely,
the dileptons produced from the pion-pion
annihilation via Vector Meson Dominance (VMD) \cite{sakurai}
as shown in Fig. 1. However, the general remarks regarding 
the effect of the Bose enhancement would be applicable to
any other model. To this end we also provide a `thumb rule' to
estimate the extent of the effect of BE at the peak of the dilepton
spectrum.

Actual estimation of the total dilepton spectra requires
a thorough understanding of the dynamics of the evolution from the very 
initial stage of the collision. In fact, many calculations, have already
been performed, incorporating various effects like collision broadening,
in-medium effects in the form of the effective mass in dense nuclear
matter motivated by the novel idea of the partial restoration of chiral
symmetry, softening of the pion dispersion relation 
 \cite{dks96,li95,chan96,koch96}.
Still a complete understanding of the data requires
further efforts in this direction. With this
emerging scenario in the background, we calculate the effect of the
nucleon loop at finite temperature together with the 
role of the Bose enhancement in the final state taking a particular
channel into account. It is seen, in this model, 
that with temperature the $\rho$- meson
mass goes down while on the other hand the $\rho$- meson decay width at
finite temperature shows an interesting behaviour as exposed below.


The importance of the $\pi\pi\rightarrow e^+e^-$ reaction in the 
context of high energy HIC has already
been addressed by many authors \cite{cassing95,wolf},
especially to explain the observed enhancement of the dilepton
yield in the low invariant mass region as reported by DLS collaboration
and most importantly in recent times in the SPS 200GeV S+Au reaction. 
Various attempts have been made to account for the data using different
models. Li, Ko and Brown  explains the data 
by taking the density dependent masses of the vector mesons \cite{li95}
where, however, Bose-enhancement effects on decay width were not taken
into account. 
Chanfray observes that medium modified pion dispersion relation is responsibile
for the excess production of the dileptons in the low mass region \cite{chan96},
Haglin shows that the other hadronic processes, in particular, $\pi\rho
\rightarrow a_1(1260)\rightarrow \pi e^+e^-$, has favourable kinematics
to populate masses between $2m_\pi$ and $m_\rho$, although no medium 
modification for the $\rho$ meson mass or decay width is considered
there \cite{hag96}. In contrast we show that the effect of the dropping
rho-meson mass is partly offset by the increase in decay width of $\rho$.

Theoretical study of $\pi^+\pi^-\rightarrow l^+l^-$ dates back to early
eighties. Domokos and Goldman were among the first to compute the
rate for this channel using relativistic kinetic theory in which
free space pion electromagnetic form factor was used and no thermal
effects on the $\rho$-meson were considered \cite{domokos}. Later, Pisarski 
in the context of chiral symmetry restoring phase transition included
some finite temperature effects on the $\rho$-meson \cite{pisarski}. 
The VMD effects,
on the other hand, have been used extensibly at finite temperature
by Gale and Kapusta, in which, the sensitivity of the dilepton yield 
on the $\pi-\rho$ dynamics in presence of a hot pion gas,
has been investigated and they concluded that
medium corrections are rather modest up to T=150 MeV \cite{gale91}. 
In the present paper, the thermal nucleon loop brings in
a modification of the rho dominated pion form factor
in the medium which is juxtaposed with the
effect of the thermal distribution of the pions. Unlike the situation
in ref. \cite{gale91} ( concentrated on $\rho\pi$ dynamics ) where the
thermal effects through the pion loop had little effect, here the
nucleon loop at finite temperature is found to affect the dilepton yield
substantially essentially because the nucleon in contrast to the pion
suffers a considerable mass reduction.  
In-medium cross-sections
are estimated considering both the effect of the thermal distribution on
the phase space and also its influence on the rho-decay widths in the
form of the Bose enhancement\cite{weldon} due to induced emission of pions in a
medium of the pion gas. 
Here a gas of pions are converted into dileptons via rho meson by the
VMD where necessary temperature dependent effects are
included. It is observed that the interaction
of the rho meson with the thermal bath causes an enhancement of the
dilepton yield in the low invariant mass region 
because of the lowering of the $\rho$- meson mass at finite temperature
while the Bose enhancement,
because of the thermal distribution of pions,
causes a reduction by increasing the rho decay widths in a hot pion gas.
Still, an overall excess yield is observed taking both the factors into
account together with the shift of the peak position towards the low
invariant mass region. To incorporate the effect of the nucleon thermal 
loop, we take recourse to the usual imaginary time formalism of 
finite temperature field theory.

The Lagrangian describing $\rho-N$ interaction is taken to be
\begin{equation}
{\cal L}_{int} = g_{\rho NN} [{\bar{N}} \gamma _\mu \tau^a 
N - \frac{\kappa}{2M}{\bar{N}} \sigma_{\mu\nu}\tau^a 
N\partial ^\nu]\rho^\mu_a 
\end{equation}							 
$\tau^a$ are the isospin Pauli matrices and N is the nucleon field. The
coupling constant and the strength of the `magnetic interaction' is determined
either from the fitting of the nucleon-nucleon interaction data 
($g_{\rho NN}=2.63,\kappa=6.1$) or from the
VMD of nucleon form factors ($g_{\rho NN}=2.71,\kappa=3.7$).

The form of the polarization tensor for the rho meson is given by 
\begin{equation}
\Pi_{\mu\nu}(q^2)=\frac{2g^2_{\rho NN}}{\beta}\sum_n\int\frac{d^3k}{(2\pi)^3}
Tr[G(\vec k+\vec q,\omega_n)\Gamma_\mu G(\vec k,\omega_n)\tilde\Gamma_\nu]
\end{equation}
where $\beta=1/kT$, $\Gamma_\mu,\tilde\Gamma_\nu$ are the appropriate
vertex factors obtained from the Lagrangian. The sum is carried out over the
Matsubara frequencies \cite{kapusta}, viz. $\omega_n=i(2n+1)\frac{\pi}{\beta}$.
This is analogous to what Blaizot discusses in the context of photon-self
energy for vacuum polarization of the QGP plasma \cite{blaiz}. 
Here, ofcourse
we have an additional contribution coming from the tensorial form of the
$\rho-N$ interaction. In the last equation
$G(\vec k,\omega_n)$ is the thermal nucleon propagator.

\begin{equation}
G(\vec k,\omega_n)=\int_0^\beta d\tau e^{\omega_n\tau}G(\vec k,\tau)
\end{equation}

\begin{equation}
G(\vec k,\tau >0)=\Lambda_+(k)\gamma_0(1-n_-(k))e^{-\epsilon_k\tau}+
\Lambda_-(k)\gamma_0n_+(k)e^{\epsilon_k\tau}
\end{equation}

\begin{equation}
G(\vec k,\tau <0)=-\Lambda_+(k)\gamma_0n_-(k)e^{-\epsilon_k\tau}-
\Lambda_-(k)\gamma_0(1-n_+(k))e^{\epsilon_k\tau}
\end{equation}

Here $n_{\pm}(k)=1/(e^{\beta\epsilon_k} \pm 1)$ and 
$\Lambda_{\pm}(k)=\frac{1}{2\epsilon_k}(\epsilon_k\pm (\alpha\cdot k +
m\gamma_0))$, the latter is the projection operator as discussed 
in Ref. \cite{blaiz} and $\epsilon_k=\sqrt{k^2+m^{\ast 2}}$, where the 
asterisk reminds us that here, instead of the free nucleon mass,
the medium modified mass is to be used.

The effective nucleon mass is calculated self-consistently
by taking only the tadpole
diagram arising out of the sigma meson exchange. 
Here also imaginary time formalism is invoked \cite{kapusta} and one subtracts 
out the divergent part as only the shift in the nucleon mass is relevant. 

The covariant rate of the dilepton production for a static thermal 
system is given by the expression
\begin{equation}
\frac{dN}{d^4x}=\int 
d^3{\tilde{p_1}}d^3{\tilde{p_2}}f(E_1)f(E_2)
|{\cal{M}}|^2(2\pi)^4\delta(l_++l_--p_1-p_2)
d^3{\tilde{l_+}}d^3{\tilde{l_-}}
\end{equation}							 
where $d^3\tilde{p_i}=\frac{d^3p_i}{2E_i(2\pi)^3}$ and $f(E_1)$, $f(E_2)$ are
the pion distribution functions. First, we perform the integration over
lepton momenta and then the integration over the initial momenta are done
taking into account 
the fact that, here, the collision is not necessarily colinear. 

The observed dilepton spectra originating from an expanding hadronic 
system (in our case a pionic gas) is obtained by convoluting the 
static (fixed temperature) emission rate with the expansion dynamics.
Relativistic hydrodynamics is a convenient tool to describe the space
time evolution of such hadronic systems \cite{landau}. In this work
we will use the Bjorken model \cite{bjorken} of boost invariant
longitudinal expansion, according to which the cooling of the system is
governed by $T(\tau)=(\tau_i/\tau)^{1/3}T_i$, where $T_i$ is the
initial temperature at the initial proper time $\tau_i$.
Within the framework of this model the invariant mass distribution
of dileptons at mid-rapidity (y=0) is given by,
\begin{equation}
\frac{dN}{dM^2dy}=\frac{3}{\pi R_A^2}(\frac{c}{4a_k}\frac{dN}{dy})^2
\frac{1}{2(2\pi)^4}\int p_Tdp_T\frac{dT}{T^7}M^2\tilde{\sigma}_i(M)K_0(M_T/T)
\end{equation}
where $R_A$ is the radius of the projectile nucleus, $c\approx 3.6$, 
$a_k=\pi^2/30$, $dN/dy$ is the
pion multiplicity, $K_0$ is the
zeroth order modified Bessel function of second kind.
$\tilde\sigma_i(M)$ is the in-medium 
cross section, and also it
is to be noted that in eq.(6) the matrix element also suffers medium
modification because of the thermal interaction as represented by the
blob in Fig. 1. 
The in medium decay width is also determined in a similar fashion by taking
the Bose enhancement factor due to the thermal distributions of the pions
and, in fact, $\Gamma_{\rho\rightarrow \pi\pi}$ appears in the denominator
of the matrix element ${\cal M}$ together with the effective nucleon mass. 
In none of these cases, however, the vertex correction is taken into account.

In Fig. 2, the variation of the effective mass of the nucleon and
the $\rho$ meson is presented to show the consistency of our calculation
with that of others, say for example, Ref. \cite{song}. One observes that
with temperature(T) the $\rho$ meson mass decreases appreciably which results
in a shift of the total dilepton yield towards the low invariant mass
region. 


The variation of the in-medium
decay width $(\Gamma_\rho)$ of the rho meson 
as a function of T which shows an interesting behaviour 
is depicted in Fig. 3.
Actually
the decay width, at finite temperature, is expected to show
a reduction as the phase
space available is less because of the reduction of the $\rho$ meson
mass in the thermal bath, while, on the other hand the presence of the
pions would cause an enhancement of the $\rho\rightarrow \pi\pi$
decay as evident from the Fig. 3. 
The rapid fall of the $\rho$-mass with increasing temperature and the
consequent reduction in $\rho$-width is dramatically offset by the 
Bose enhancement so that upto a temperature of 160 MeV, 
the width remains in the region $\sim$155 to 161 MeV.
This particular feature of the
$\rho$ decay width has not been remarked upon earlier. 
The $\rho\pi\pi$ coupling constant,
is estimated from the decay width of the rho meson in free
space and is given by $g^2_{\rho\pi\pi}/4\pi\sim$ 2.9. We also
note from Fig. 2 that the nucleon effective mass, while showing a reduction
at finite T, is still large compared to the rho meson mass and
therefore the nucleon-antinucleon channel would still be closed and
have no effect on the decay width.


To bring the effect of Bose enhancement into bold relief, 
we discuss the effect of the nucleon thermal loop on the 
dilepton spectra with and without the Bose enhancement factor
included for the $\rho\rightarrow\pi\pi$ channel. 
In our example we have taken $R_A=4.6$ fm, $dN/dy=225$, $T_i=185$ MeV,
$T_F=130$ MeV, the results are plotted in fig.4,
where shift of the peak position towards a low invariant mass is observed 
together with an enhancement of the total yield. It is to be noted that
inclusion of the Bose enhancement suppresses the yield substantially 
by increasing the decay width as
expected from Fig. 3. Quantitatively, a suppression $\sim 30\%$ 
of the total dilepton yield is
observed when the effect of BE on decay width is considered compared
to the results 
without BE effects everything else remaining the same.
However, still an excess production of the dileptons
at finite temperature is expected once we include the effect of the thermal
loop. Our findings without the effect of BE on decay width 
are consistant with what Li, Ko and Brown observes taking
into account the matter induced modification of the $\rho$ meson mass.
It may be noted that upto a temperature(T) $\sim 100-150$ MeV, 
the BE is dominant as evident from Fig. 3 beyond which the mass 
modification takes over causing an over all  reduction of the decay width
indicating the sensitivity of the total dilepton yield on the initial
temperature of the hadronic phase.
To  bring this effect into clear focus we  also  present
the spectra for $T_i=165$
MeV, keeping all the other parameters same as before. The results are
displayed in fig.5. 


The Bose enhancement factor for the rho width due to induced emission
of pions is clearly given by 
$(1 + e^{-\beta m_\rho/2})^2$ 
as each pion carries the energy $m_\rho/2$.
Thus the square modulus of the vector meson dominated pion form factor
at the $\rho$-peak $(\frac{m_\rho^2}{\Gamma ^2})$ receives a reduction
factor of $(1+e^{-\beta m_\rho/2})^{-4}$ and accordingly this would reflect
itself on the lepton yield from this channel. However, since the basic
process has to be integrated over the space-time evolution and the
concomittant thermal history incorporated in the above reduction factor, 
an effective temperature $(T_{eff})$ needs to be inserted. In order to 
determine $T_{eff}$, the peak in the invariant mass of the total yield
is found and in the given model the temperature at which the vector
meson mass achieves the value $m_{peak}$ gives us an estimate of 
$T_{eff}$. {\em {Thus the effective reduction factor may be estimated to be 
$(1 + e^{- m_\rho/{2kT_{eff}}})^{-4}$. It is easily checked that this 
roughly agrees with our detailed determination of this value. It should
be noted that in a case where the vector meson ( say $\omega$ ),
dominantly decays into three pions the suppression due to Bose enhancement
will be the sixth power of the relevant analogous factor}}.

In conclusion, we observe that the thermal nucleon loop and the thermal
distribution of the pion gas affect the dilepton yield in the low 
invariant mass region quite appreciably and should therefore be included
in the dynamical calculation to move towards a more complete understanding
of the dilepton production in high energy heavy ion collision. 
The main result of the present analysis is that the effect of BE on the
widths of vector mesons with dominant pionic decay channels can have
substantial influence on the dilepton yield depending on the initial
temperature.  Studies are in progress to investigate these effects for 
other hadronic channels which may contribute to the production of leptons. 

We gratefully acknowledge useful discussions with J.P. Blaizot (SACLAY)
K. Haglin (MSU), R. Bhalerao (TIFR), D. K. Srivastava (VECC) and 
K. Mukherjee (SINP). One of the author (BS) acknowledges
stimulating discussions with C. M. Ko, G. E. Brown and M. Prakash.

\newpage
\centerline{Figure captions}
\begin{enumerate}
\item Feynman diagram for dilepton production via Vector Meson Dominance
from pion-pion annihilation where the blob represents the interaction of
the $\rho$ meson with the thermal bath.

\item{ }
Effective masses are depicted as a function of temperature both for
nucleon and rho meson taking the effect of thermal loops into account.

\item{ }
The variation of the decay width with temperature for three different
cases are shown. The value of the free decay width is 
151 MeV. Solid, dashed and dotted curves represent decay width using
the free $\rho$ mass with BE, medium modified $\rho$ width 
with and without BE respectively.

\item{ }

Effect of the thermal bath on the dilepton yield for $T_i=185 MeV$ and
$T_F= 130$ MeV. Solid, dashed,  
and dotted represent contribution with 
free (with out medium effect),
medium-modified without BE factor and medium-modified with BE factor 
respectively.

\item{ }

Same as Fig.4 for $T_i=165 MeV$.

\end{enumerate}

\newpage
\begin{thebibliography}{99}

\bm {gale} C. Gale and J. Kapusta, Phys. Rev {\bf C 35}(1987)2107
{\bf C 38}(1988)2657

\bm {xia} L.H. Xia, C.M. Ko, L. Xiong and J.Q. Wu, Nucl Phys 
{\bf A 485}(1988)721

\bm{dks96} D.K. Srivastava, B. Sinha and C. Gale  Phys. Rev. C{\bf 53}
(1996)R567

\bm {sakurai} J.J. Sakurai, Currents and Mesons (Univ. of Chicago Press,
Chicago, 1969).

\bm {li95} G.Q. Li, C.M. Ko and G.E. Brown 
Phys. Rev Lett {\bf 75} (1995)4007

\bm {chan96} G. Chanfray, R. Rapp and J. Wambach 
Phys. Rev Lett {\bf 76} (1996)368

\bm {koch96} V. Koch and C. Song, Pre-print LBL-38619/UC-413

\bm {cassing95} W. Cassing, W. Ehehalt and C.M. Ko 
Phys. Lett {\bf B 363}(1995)35

\bm {wolf} Gy. Wolf, W. Cassing and U. Mosel, Nucl Phys 
{\bf A 552}(1993)549

\bm {hag96} K.L. Haglin, Phys. Rev. {\bf C 53}(1996)R2606

\bm {domokos} G. Domokos and J.I. Goldman, Phys. Rev. {\bf D 23}(1981)203

\bm {pisarski} R.D. Pisarski, Phys. Lett. {\bf B110}(1982)155

\bm {gale91} C. Gale and J.I. Kapusta, Nucl. Phys. {\bf B357}(1991)65

\bm {weldon} H.A. Weldon, Phys. Rev. {\bf D 28}(1983)2007

\bm {blaiz} J.P. Blaizot, SACLAY-T94/044

\bm {hag93} K.L. Haglin, C. Gale and V. Emel'yanov,
Phys. Rev. {\bf D 47}(1993)973

\bm {kapusta} J.I. Kapusta, Finite Temperature Field Theory
( Cambridge University Press, Cambridge, 1989).

\bm {song} C. Song, P.W. Xia and C.M. Ko, Phys. Rev. {\bf C 52}(1995)408

\bibitem {landau} D. ter Haar (ed.), Collected Papers of L. D. Landau,
(Gordon and Breach, London, 1965)p.665.

\bibitem {bjorken} J. D. Bjorken, Phys. Rev. {\bf D27} (1983)140.

\end{thebibliography}
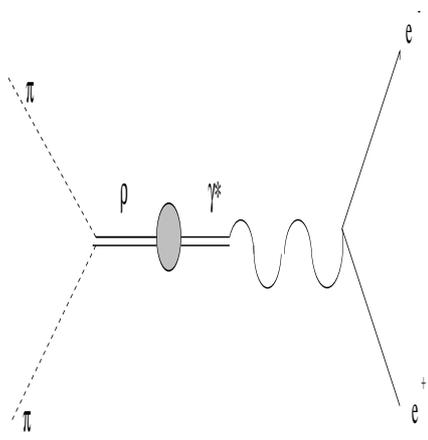
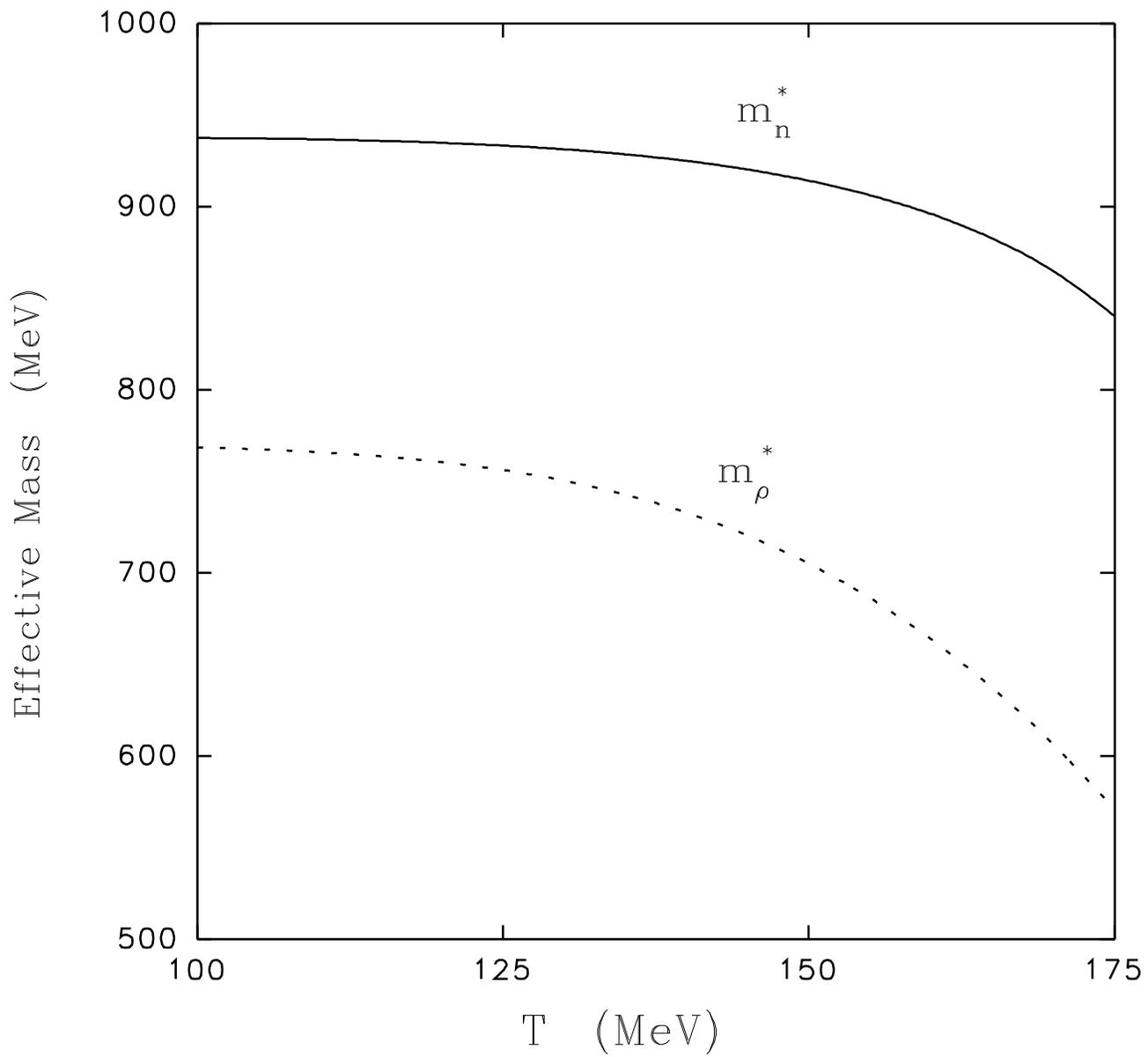
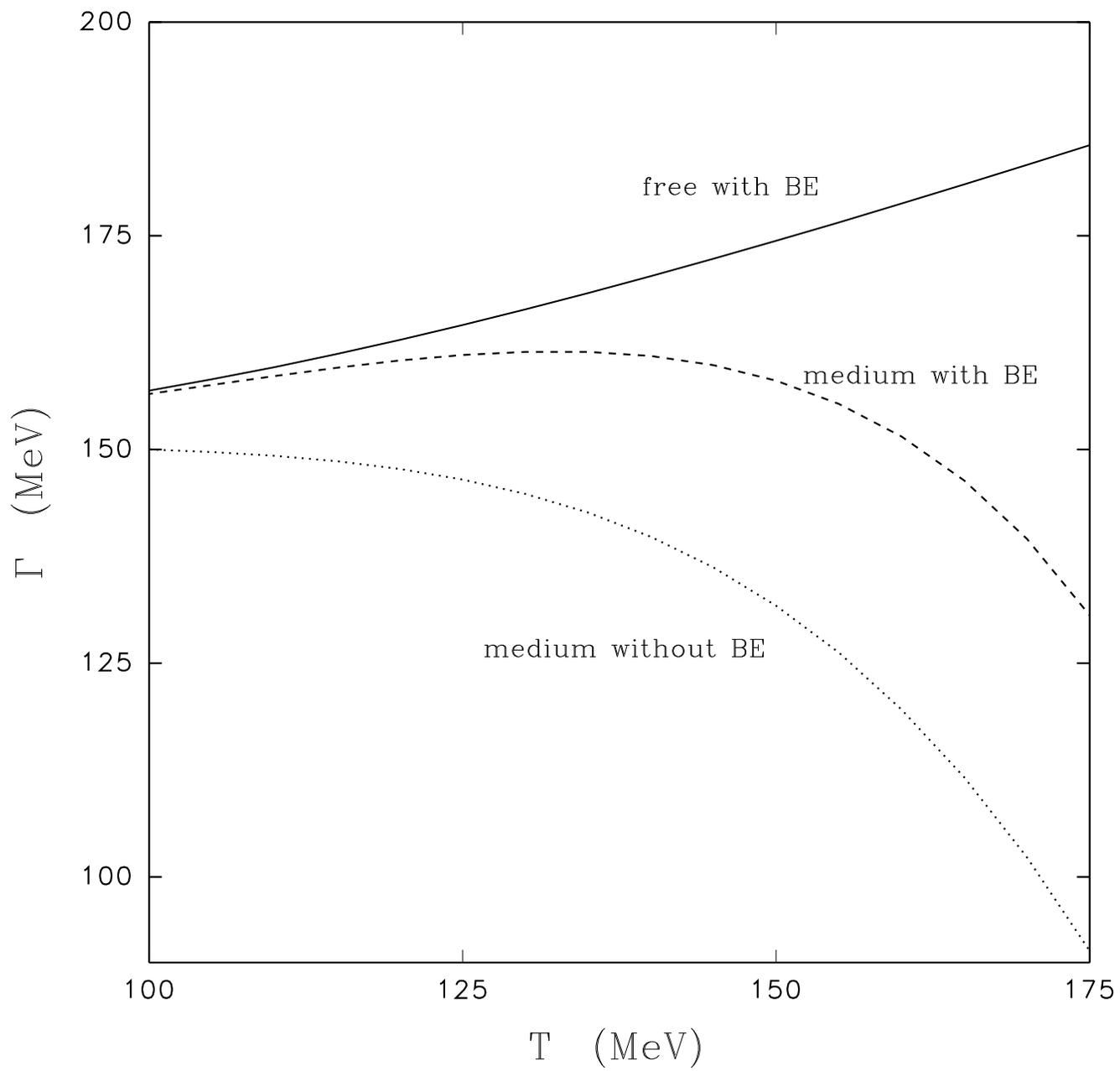
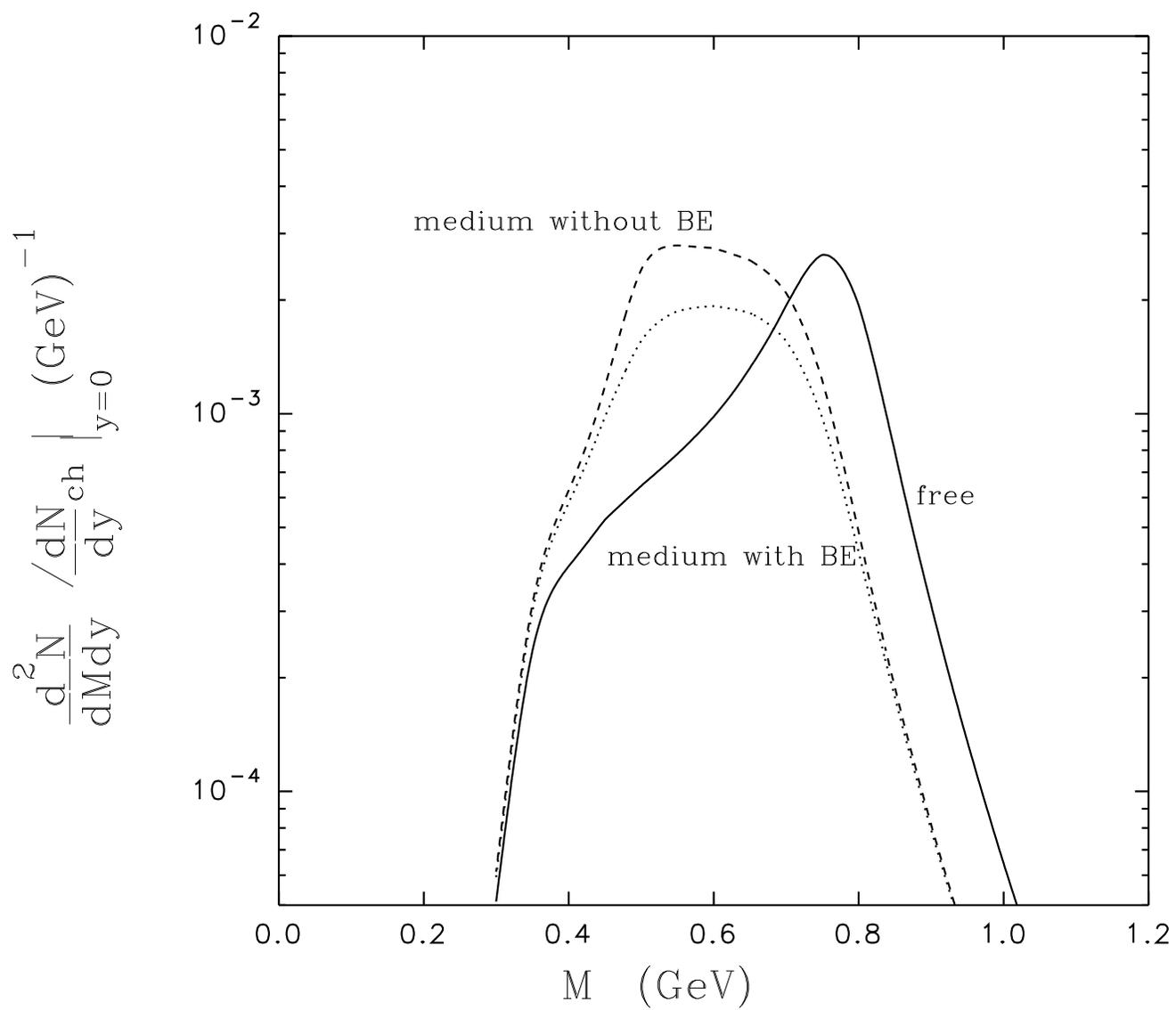
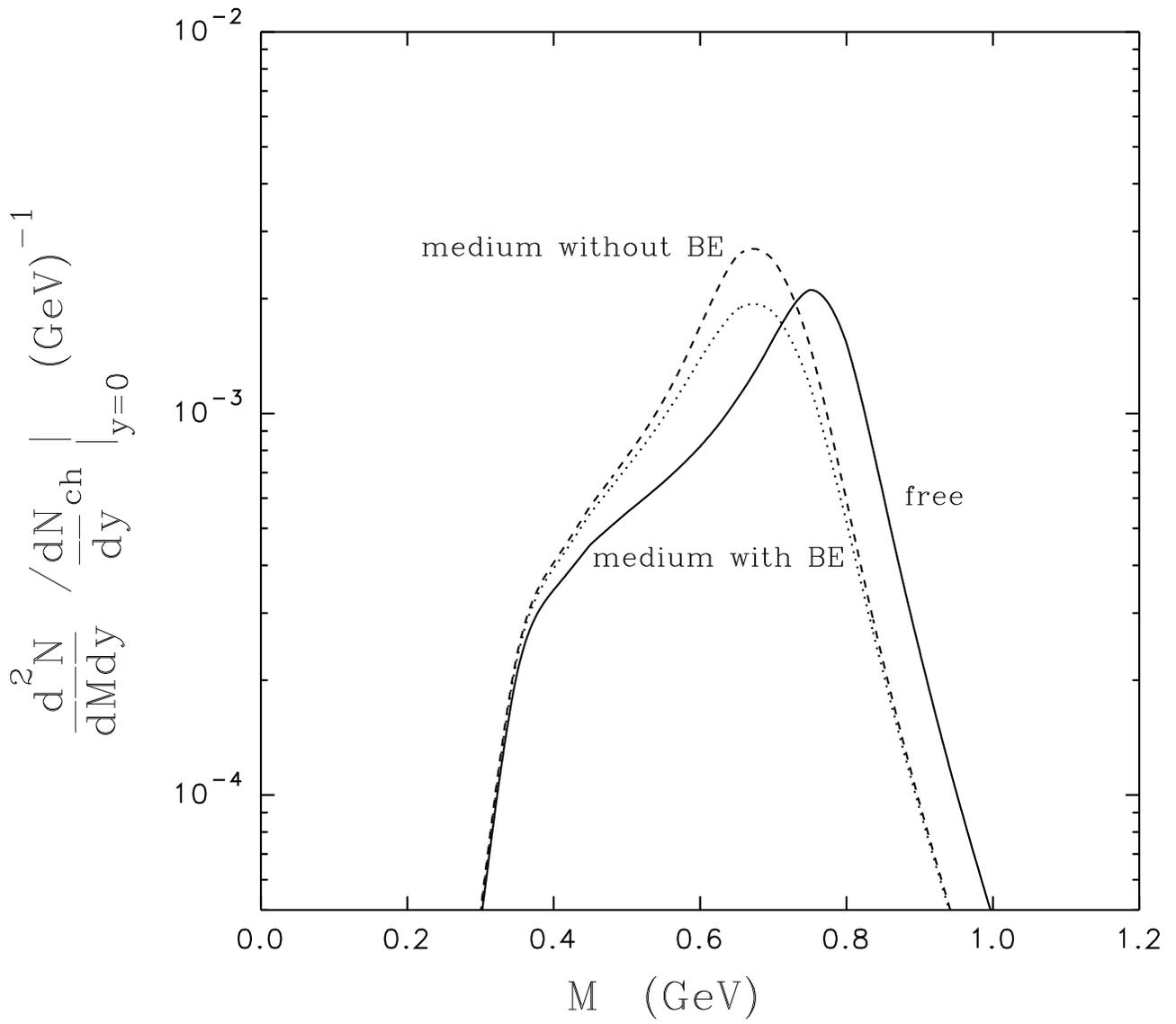
\end{document}